# A First Principle Study on Iron Substituted LiNi (BO₃) to use as Cathode Material for Li-ion Batteries

Anu Maria Augustine[1,2], Vishnu Sudarsanan[1,2], Geo Sunny[1] and P. Ravindran[1,2,3,4 a)]

[1]*Department of Physics, Central University of Tamil Nadu, Thiruvarur, Tamil Nadu, 610101, India*
[2]*Simulation Center for Atomic and Nanoscale MATerials (SCANMAT), Central University of Tamil Nadu, Thiruvarur, Tamil Nadu, 610101, India*
[3]*Department of Materials Science, Central University of Tamil Nadu, Thiruvarur, Tamil Nadu, 610101, India*
[4]*Center for Material Science and Nanotechnology and Department of Chemistry, University of Oslo, Box 1033 Blindern, N-0315 Oslo, Norway*

a)Corresponding author: raviphy@cutn.ac.in

**Abstract.** In this work, the structural stability and the electronic properties of $LiNiBO_3$ and $LiFe_xNi_{(1-x)}BO_3$ are studied using first principle calculations based on density functional theory. The calculated structural parameters are in good agreement with the available theoretical data. The most stable phases of the Fe substituted systems are predicted from the formation energy hull generated using the cluster expansion method. The 66% of Fe substitution at the Ni site gives the most stable structure among all the Fe substituted systems. The bonding mechanisms of the considered systems are discussed based on the density of states (DOS) and charge density plot. The detailed analysis of the stability, electronic structure, and the bonding mechanisms suggests that the systems can be a promising cathode material for Li ion battery applications.

## INTRODUCTION

Improving the battery technology is critical for the development of various fields ranging from the portable vehicles to cell phones, laptop computers and power tools industry. Among the various types of batteries, Li ion battery technology is the prominent one with high energy density and greater operational lifetime. [1] The Li ion battery technology is manifold in the chemistry involved with the use of different cathode materials and these cathode materials are one of the main characteristic component which defines the performance of the battery. [2] Earlier works on $LiNiBO_3$ suggest that it is a promising cathode material with energy density 1215 W/l and average voltage 4.71 eV. The study of electronic properties of the material gives more insight into its potential to be a good cathode material in Li ion battery. So here we present the study of phase stability and electronic properties of the polyanionic compound $LiNi(BO_3)$ [3] and the Fe substituted $LiNi(BO_3)$ systems.

## COMPUTATIONAL DETAILS

The total energies and the optimized structural details were obtained using the density functional theory within the generalized gradient approximation of Perdew-Burke-Ernzerhof [4] for the exchange-correlation functional with projector augmented wave potentials. We used the Vienna *ab initio* Simulation Package (VASP) [5] which implements the periodic boundary conditions and the plane wave basis sets. Force as well as stress minimization were done to obtain the structural optimization. Plane wave basis with a cut-off kinetic energy of 520 eV is used and the total energies were calculated by integration over a Monkhorst-Pack mesh of **k**-points in the Brillouin zone with sufficient **k**-point density.

We used the Cluster Assisted Statistical Method (CASM) code [6] which functions using the group theoretical techniques to automate the construction and parameterization of effective Hamiltonians. The formation energy for all compounds discussed here are calculated with the help of these effective Hamiltonians. In order to account for the magnetic polarization we have made all the calculations in the ferromagnetic configuration. The correlation effects that influence the structural and electronic properties we have made the calculations using GGA+U method with U value of 6.2 eV for the Ni-d electrons and 5.3 for the Fe-d electrons.

## STRUCTURAL DETAILS

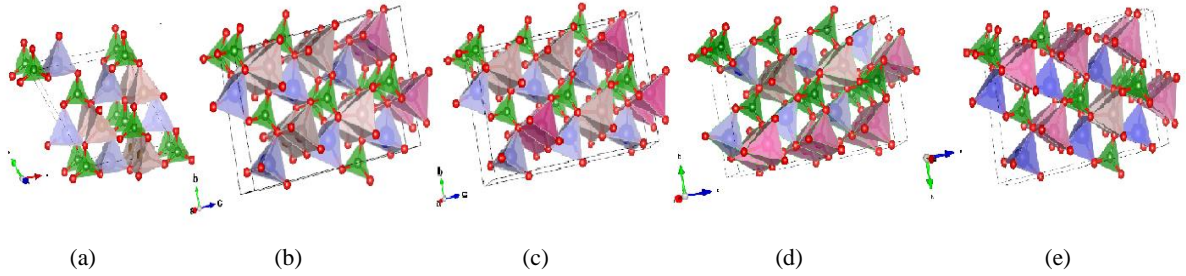

(a)  (b)  (c)  (d)  (e)

**FIGURE 1.** Ground state crystal structure of (a) $LiNiBO_3$, (b) $LiFe_{0.33}Ni_{0.66}(BO_3)$, (c) $LiFe_{0.55}Ni_{0.45}(BO_3)$, (d) $LiFe_{0.66}Ni_{0.34}(BO_3)$ and (e) $LiFe_{0.85}Ni_{0.15}(BO_3)$. Li, Ni, Fe, B, and O are represented in violet, brown, magenta, green, and red color respectively.

In the present study we have considered the hexagonal form of $LiNiBO_3$ with space group $P\bar{6}$. The Fe atoms are substituted at Ni site in different concentrations. The optimized structure of $LiFe_xNi_{(1-x)}BO_3$ with various concentration of Fe is shown in Figure 1 and the optimized structural parameters are depicted in Table 1. The calculated lattice parameters of $LiNiBO_3$ is found to be in good agreement with previous works based on similar calculation [3].

**Table 1** Optimized structural parameters for pristine $LiNiBO_3$ and Fe substituted $LiNiBO_3$.

| Compound | Lattice parameters (Å) a, b, c | Volume (Å$^3$) |
|---|---|---|
| $LiNi(BO_3)$ | 8.00544, 8.00542, 3.05228 | 169.4009 |
|  | (8.073, 8.072, 3.056) [3] | 172.482 |
| $LiFe_{0.33}Ni_{0.66}(BO_3)$ | 3.09123, 8.09080, 13.99789 | 350.0947 |
| $LiFe_{0.55}Ni_{0.45}(BO_3)$ | 3.09762, 8.12582, 14.04129 | 353.4251 |
| $LiFe_{0.66}Ni_{0.34}(BO_3)$ | 3.10628, 8.11095, 14.10397 | 355.3472 |
| $LiFe_{0.85}Ni_{0.15}(BO_3)$ | 3.13848, 8.12808, 14.08386 | 359.2758 |

## RESULTS AND DISCUSSION

In order to determine the most stable configuration with various Fe concentration we have plotted the formation energy hull as given in Figure 2. This figure shows the formation energies of 203 different configuration for 6 different compositions. We have used CASM code to find the formation energies for different configurations of same composition. The configurations which fall on the convex hull (line which connects the most stable configurations in Figure (2)) are considered for further studies. From this figure it can be noted that the Fe substituted $LiNiBO_3$ shows most stable configuration when x=0.66. The compounds with partial substitution of Fe at Ni site show more stability than the parent compound. Whereas the full substitution of Fe at Ni site leads to highly unstable structure. The partial substitution of Fe atom can improve the kinetics of Li intercalation due the stability of Fe atom in +3 as well as +2 oxidation states compared to Ni atom. So it can be expected that the system with partial Fe substitutions are more preferable as a cathode material for Li ion battery. The present study on formation energy of $LiFe_xNi_{1-x}BO_3$ systems reveals its stabile nature and confirm its practicability.

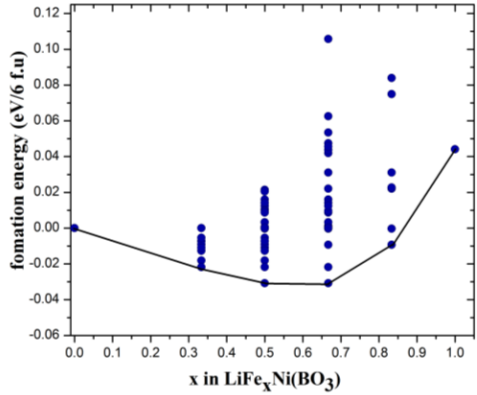

**FIGURE 2.** (a) Formation energy hull of LiFe$_x$Ni$_{1-x}$BO$_3$.

To understand the basic features of chemical bonding the total density of states of LiNiBO$_3$ and LiFe$_x$Ni$_{1-x}$BO$_3$ compounds are calculated at the equilibrium geometries and shown in Figure 3(a). The non-vanishing electronic states at Fermi level are responsible for the metallic nature of all the considered systems. Even though, the positions of peaks in DOS is not highly altered by the Fe substitution, the intensity of the DOS is changed due to the lower number of electrons in Fe atom. For a better analysis we have decomposed the total density of states of LiFe$_{0.66}$Ni$_{1-0.66}$ BO$_3$ into site projected contributions in Figure 3(b). From this figure transition metals are found to be the major contributors of the states at the Fermi level. The DOS at the Li site is negligibly small indicating that most of its electrons are transferred to the oxygen sites and hence there is an ionic bonding between Li with the host lattice. However, there is finite distribution of electronic state in the energy range between -6 to -2 eV and their energetic degenerate state with O indicating the presence of finite covalency. It can also be noted that the depletion of DOS at valance band of B reflects its ionic nature. We can also say that the DOS at higher energy ranges much above the Fermi level is derived from the unoccupied electronic states of Li and B.

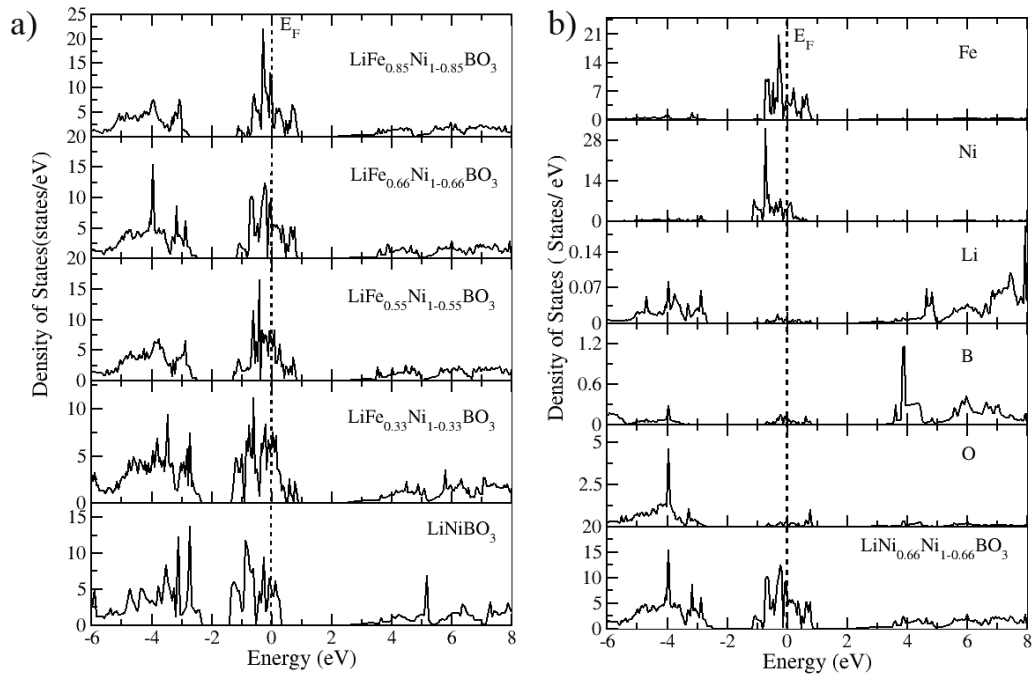

**FIGURE 3.** The total DOS of (a) LiNiBO$_3$ and Fe substituted LiNiBO$_3$ and (b) partial DOS of LiFe$_{0.66}$Ni$_{1-0.66}$BO$_3$.

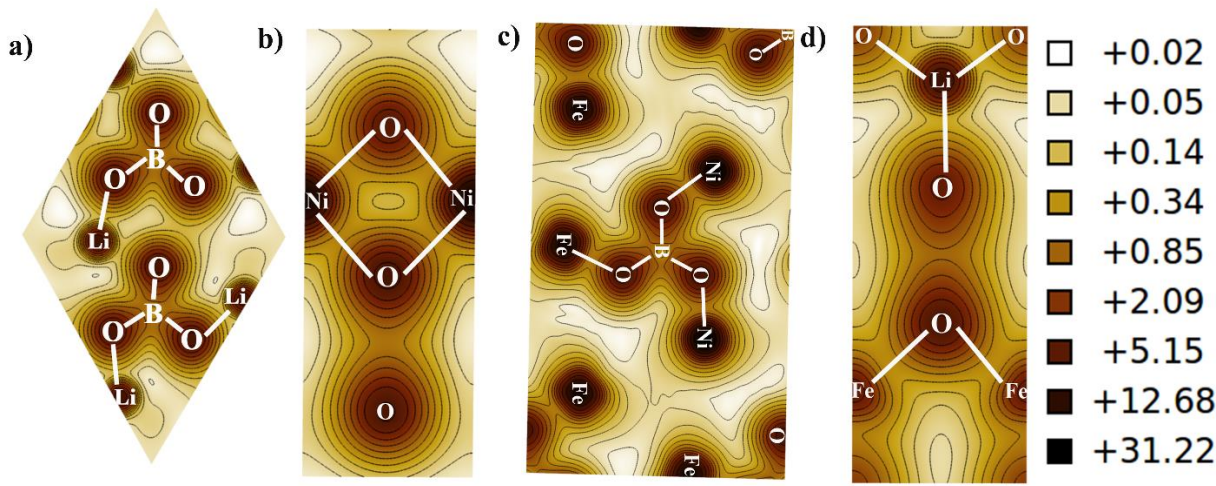

**FIGURE 4.** Charge density distribution in LiNiBO$_3$ (a and b) and LiFe$_{0.66}$Ni$_{1-0.66}$BO$_3$ (c and d) for the planes with representative atoms.

More detailed understanding of the bonding mechanism can be obtained from charge density analysis. The contour map of the charge density along characteristic planes are depicted in the Figure 4. From the figure, B-O bond can be characterized as having highly ionic nature due to charge transfer from B to O as discussed in the DOS analysis and hence no noticeable charge is present at the B sites. The electrons at the Li site is isotropic distribution and also negligible small charge present between Li and O indicating ionic bonding between Li and O. The bond between Ni and O as well as between the oxygen atoms are found to be covalent in nature due to the anisotropic distribution of charges and their presence between the corresponding atoms. The charge density analysis of the Fe substituted system also shows same kind of bonding mechanism between B-O, Li-O, and O-O as in LiNiBO$_3$. The charge density distribution between Fe and O is almost similar to that for Ni and O in the parent compound.

## CONCLUSIONS

In summary, the structural stability, the electronic properties and the bonding mechanism in LiNiBO$_3$ and LiFe$_x$Ni$_{(1-x)}$BO$_3$ have been studied using density functional theory within GGA approximation. The formation energy analysis predicts that LiFe$_{0.66}$Ni$_{1-0.66}$(BO$_3$) is the most stable phase among the Fe substituted LiNiBO$_3$ systems and also it shows more stability compared to the parent system. The DOS and the bonding analysis reveals that the electronic structure of LiNi(BO$_3$) is not drastically influenced by the partial substitution of Fe at Ni site. From all these analysis it can be concluded that Fe substituted LiNi(BO$_3$) with composition LiFe$_x$Ni$_{(1-x)}$BO$_3$ is a potential cathode material for Li-ion batteries.

## ACKNOWLEDGMENTS

The authors are grateful to the DST, India for the funding support via Grant No. SR/NM/NS-1123/2013 and the Research Council of Norway for computing time on the Norwegian supercomputer facilities.